\documentclass[aps,twocolumn,showpacs,floatfix]{revtex4}
\usepackage{amsmath}
\usepackage{amsfonts}
\usepackage{amssymb}
\usepackage{graphicx}
\begin{document}
\title{Electromagnetically Induced Transparency with Quantized Fields in Optocavity Mechanics}
\author{Sumei Huang and G. S. Agarwal}
\affiliation{Department of Physics, Oklahoma State University,
Stillwater, Oklahoma 74078, USA}
\date{\today}
\begin{abstract}
We report electromagnetically induced transparency using quantized fields in optomechanical systems. The weak probe field is a narrow band squeezed field. We present a homodyne detection of EIT in the output quantum field. We find that the EIT dip exists even though the photon number in the squeezed vacuum is at the single photon level. The EIT with quantized fields can be seen even at temperatures of the order of 100 mK paving the way for using optomechanical systems as memory elements.

\end{abstract}
\pacs{42.50.Gy, 42.50.Wk, 03.65.Ta}
\maketitle
\section{Introduction}
The interaction of a nano mechanical system via the radiation pressure \cite{Meystre,Tombesi} is like a three wave interaction in nonlinear optics \cite{Boyd}. This interaction can lead to processes like upconversion for example a photon of frequency $\omega_{c}$ can be converted into a photon of frequency $\omega_{p}=\omega_{c}+\omega_{m}$ where $\omega_{m}$ is the frequency of the mechanical oscillator. Such upconversion processes have been useful in cooling the nano mechanical systems \cite{Gigan,Cohadon,Schliesser,Wang}. In a previous paper \cite{Agarwal} we showed how such upconversion processes can lead to electromagnetically induced transparency in optomechanical systems. The EIT in such systems turned out to share many of the features of EIT in atomic vapors. The EIT in optomechanical systems has been seen experimentally \cite{Weis,Vahala,Painter}. Traditionally almost all EIT experiments in atomic systems and other systems have been done with coherent pump and probe fields \cite{Harris1,Harris2,Scully}. Akamatsu et al \cite{Akamatsu} did the very first experiment on EIT using the squeezed light in atomic vapors. They essentially reported that squeezing of the probe is not graded much by the quantum noise of the medium under EIT conditions. Following this a number of other experiments \cite{Kozuma,Arikawa} on EIT using quantized fields were reported. The EIT with quantized fields is very significant in storage of fields at single photon level \cite{Lukin, Lvovsky, Lvovsky09, Kozuma08}.

In this paper we examine EIT in optomechanical systems using the quantized fields. In optomechanical systems the noise is added both by the resonator's noise as well as the noise of the mechanical system. We would find conditions when perfect EIT of the quantized field results. We would study how the temperature of the mechanical system can degrade the EIT. We would present detailed results for designs of nano mechanical systems as used in refs \cite{Weis,Aspelmeyer}. We find that certain designs of nano mechanical systems are good even at temperatures of the order of 100mK. Thus such systems would be quite useful as optical memories at single photon level.  The results that we present can be extended to reactive case \cite{Li,Sumei1,Sumei2}.

The organization of the paper is as follows. In Sec. II, we describe the model, derive the equations of the motion for the system, and obtain the steady-state mean values. In Sec. III, we show how to detect EIT with quantized fields, we present a homodyne detection and obtain the relevant spectrum. In Sec. IV, we discuss the impact of the coupling field on the homodyne spectrum of the output field, and show the existence of the EIT in the homodyne spectrum of the quantized field at the output.
\begin{figure}[htp]
 \scalebox{0.8}{\includegraphics{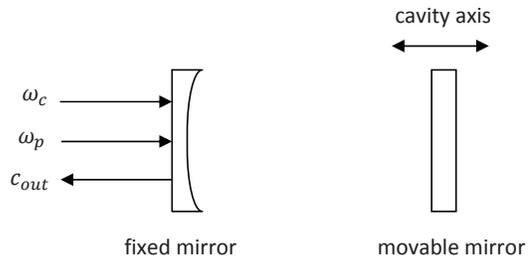}}
 \caption{\label{Fig1} Sketch of the studied system.
A coherent coupling field at frequency $\omega_{c}$ and a squeezed vacuum at
frequency $\omega_{p}$ enter the cavity through the partially
transmitting mirror.}
 \end{figure}
\section{Model}
The model we are going to consider has been discussed in detail before \cite{Jahne,Sumeisq} and is sketched in Fig. 1. The cavity consists of a fixed mirror and a movable mirror separated by a distance $L$. The fixed mirror is partial transmitting, while the movable mirror is 100\% reflecting. The cavity is driven by a strong coupling field at frequency $\omega_{c}$. A quantized weak probe field in a squeezed vacuum state at frequency $\omega_{p}$ is injected into the cavity through the fixed mirror. The movable mirror interacts with the cavity field through the radiation pressure. The movable mirror is modeled as a harmonic oscillator with mass $m$, frequency $\omega_{m}$, and decay rate $\gamma_{m}$. Moreover, the movable mirror and its environment are in thermal equilibrium at a low temperature $T$.

In such a system, the coupling between the movable mirror and the cavity field is dispersive so that the frequency $\omega_{0}(q)$ of the cavity field depends on the displacement $q$ of the movable mirror: $\omega_{0}(q)=n\pi c/(L+q)$, where $c$ is the light speed in vacuum, and $n$ is the mode number in the cavity. For $q\ll L$, we can expand $\omega_{0}(q)$ to the first order of $q$, thus we have $\omega_{0}(q)\approx\omega_{0}(0)+\frac{\partial\omega_{0}(q)}{\partial q}q\approx\omega_{0}-\frac{\omega_{0}}{L}q$, where we write $\omega_{0}(0)$ as $\omega_{0}$.

Let $c$ ($c^{\dag}$) be the annihilation (creation) operators for the cavity field, $Q$ and $P$ be the dimensionless operators for the position and momentum of the movable mirror with $Q=\sqrt{\frac{2m\omega_{m}}{\hbar}}q$ and $P=\sqrt{\frac{2}{m\hbar\omega_{m}}}p$. Note that the commutation relation for $Q$ and $P$ is $[Q,P]=2i$.  In a frame rotating at the frequency $\omega_{c}$ of the coupling field, the Hamiltonian for the system is
\begin{eqnarray}\label{1}
H&=&\hbar(\omega_{0}-\omega_{c})c^{\dag}c-\hbar g c^{\dag}c
Q+\frac{\hbar\omega_{m}}{4}(Q^2+P^2)\nonumber\\
& &+i\hbar\varepsilon(c^{\dag}-c),
\end{eqnarray}
In the above equation, the parameter $g=(\omega_{c}/L)\sqrt{\hbar/(2m\omega_{m})}$ is the coupling strength between the cavity field and the movable mirror, where we assume $\omega_{0}\simeq\omega_{c}$. The parameter $\varepsilon$ is the real amplitude of the coupling field, depending on its power $\wp$ by $\varepsilon=\sqrt{\frac{2\kappa\wp}{\hbar\omega_{c}}}$, where $\kappa$ is the photon loss rate due to the transmission of the fixed mirror.

The time evolution of the total system is obtained from Hamiltonian (\ref{1}) by deriving the Heisenberg
equations of motion and adding the damping and noise terms. The basic equations are given by
\begin{equation}\label{2}
\begin{array}{lcl}
\dot{Q}=\omega_{m}P,\vspace*{.1in}\\
\dot{P}=2g n_{c}-\omega_{m}Q-\gamma_{m}P+\xi,\vspace*{.1in}\\
\dot{c}=i(\omega_{c}-\omega_{0}+g
Q)c+\varepsilon-\kappa c+\sqrt{2
\kappa}c_{in},\vspace*{.1in}\\
\dot{c}^{\dag}=-i(\omega_{c}-\omega_{0}+g
Q)c^{\dag}+\varepsilon-\kappa
c^{\dag}+\sqrt{2\kappa}c_{in}^{\dag}.
\end{array}
\end{equation}
Here we have introduced the thermal Langevin force $\xi$ with vanishing mean value, resulting
from the coupling of the movable mirror to the environment. The Langevin force $\xi$ has the correlation function in the frequency domain:
\begin{eqnarray}\label{3}
\langle\xi(\omega)\xi(\Omega)\rangle=4\pi\gamma_{m}\frac{\omega
}{\omega_{m}}\left[1+\coth\left(\frac{\hbar\omega}{2K_B T}\right)\right]\delta(\omega+\Omega),\nonumber\\
\end{eqnarray}
where $K_{B}$ is the Boltzmann constant. Throughout this paper the following Fourier relations are used
\begin{eqnarray}\label{4}
f(t)&=&\frac{1}{2\pi}\int_{-\infty}^{+\infty}f(\omega)e^{-i\omega t}
d\omega,\nonumber\\
f^{\dag}(t)&=&\frac{1}{2\pi}\int_{-\infty}^{+\infty}f^{\dag}(-\omega)e^{-i\omega
t} d\omega,
\end{eqnarray}
where $f^{\dag}(-\omega)=[f(-\omega)]^{\dag}$. The $c_{in}$ represents the input quantum field which is centered around the frequency $\omega_{p}=\omega_{c}+\omega_{m}$ with a finite bandwidth $\Gamma$. The quantized field has the following nonvanishing correlation functions,
\begin{eqnarray}\label{5}
\langle c_{in}(\omega)
c_{in}(\Omega)\rangle=2\pi \displaystyle\frac{M\Gamma^{2}}{\Gamma^{2}+(\omega-\omega_{m})^2}\delta(\omega+\Omega-2\omega_{m}),\nonumber\\
\langle c_{in}(\omega)
c_{in}^{\dag}(-\Omega)\rangle=2\pi\left[\displaystyle\frac{N\Gamma^{2}}{\Gamma^{2}+(\omega-\omega_{m})^2}+1\right]\delta(\omega+\Omega),\nonumber\\
\end{eqnarray}
where $N$ is the photon number in the squeezed vacuum, and $M=\sqrt{N(N+1)}$. The antinormally ordered term has a broad band contribution coming from vacuum noise. Note that by setting $M=0$, we would obtain standard phase independent quantum field with mean number of photons $\frac{N\Gamma^{2}}{\Gamma^{2}+(\omega-\omega_{m})^2}$ around the frequency $\omega=\omega_{m}$.

The mean values at steady state can be obtained from Eq. (\ref{2}) by setting all time derivatives to zero. These are found to be
\begin{equation}\label{6}
P_{s}=0,\hspace{.02in}Q_{s}=\frac{2g |c_{s}|^{2}}{\omega_{m}},\hspace{.02in}c_{s}=\frac{\varepsilon}{\kappa+i\Delta},
\end{equation}
where
\begin{equation}\label{7}
\Delta=\omega_{0}-\omega_{c}-g Q_{s}
\end{equation}
is the effective cavity detuning.

\section{The output field and its measurement}

\begin{figure}[htp]
 \scalebox{0.8}{\includegraphics{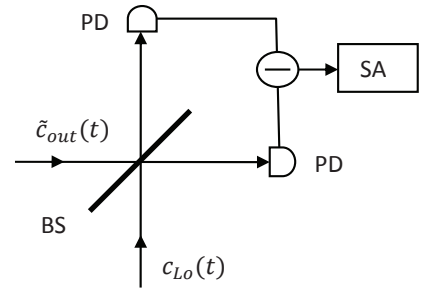}}
 \caption{\label{Fig2} Sketch of the measurement of the output field.
The output field $\tilde{c}_{out}(t)$ is mixed with a strong local field $c_{Lo}(t)$ centered around the probe frequency $\omega_{p}$ at a beam splitter, where $\tilde{c}_{out}(t)$ is defined as the sum of the output field $c_{out}(t)$ from the cavity and the input quantized field $c_{in}(t)$. BS: 50/50 beam splitter, PD: photodetector, SA: spectrum analyzer.}
 \end{figure}

The output field is a quantum field, it contains many Fourier components. Since the quantized input field is centered around $\omega_{p}=\omega_{c}+\omega_{m}$, the interesting component of the output field is near the probe frequency $\omega_{p}$. So we mix the output field $\tilde{c}_{out}(t)$ with a strong local field $c_{Lo}(t)$ centered around the probe frequency $\omega_{p}$ at a 50/50 beam splitter, as shown in Fig. 2. In a frame rotating at the frequency $\omega_{c}$, $c_{Lo}(t)=c_{Lo}e^{-i\delta_{0}t}$, where $\delta_{0}=\omega_{p}-\omega_{c}$. The difference between the output signals from the two photodetectors is sent to the spectrum analyzer, and the output signal from the spectrum analyzer depends on the phase of $c_{Lo}$. If $c_{Lo}$ is real, the homodyne spectrum $X(\omega)$ of the output field measured by the spectrum analyzer is given by
\begin{eqnarray}\label{8}
& &\langle[c_{Lo}^{*}(t)\tilde{c}_{out}(t)+c.c.][c_{Lo}^{*}(t')\tilde{c}_{out}(t')+c.c.]\rangle\nonumber\\
&=&\frac{c_{Lo}^{2}}{2\pi}\int\,d\omega e^{-i\omega(t-t')}X(\omega).\
\end{eqnarray}
Thus in our investigations of EIT with quantized fields, $X(\omega)$ is the quantity of interest.

In order to study the EIT effect in the homodyne spectrum $X(\omega)$ of the output field, we will calculate the fluctuations of the output field. The steady state part would not contribute as it is at the frequency of the coupling field. We assume the photon number in the cavity is large enough so that each operator can be written as a linear sum of the steady-state mean value and a small fluctuation, which yields
\begin{equation}\label{9}
Q=Q_{s}+\delta Q,\hspace*{.1in}P=P_{s}+\delta P,\hspace*{.1in}c=c_{s}+\delta c,
\end{equation}
where $\delta Q$, $\delta P$, and $\delta c$ are the small fluctuations around the steady state. By substituting Eq. (\ref{9}) into Eq. (\ref{2}), one can arrive at the linearized equations for the
fluctuation operators. Further we transform the linearized equations into the frequency domain by Eq. (\ref{4}) and solve it, we can obtain the fluctuations $\delta c(\omega)$ of the cavity field. Then using the input-output relation $c_{out}(\omega)=\sqrt{2\kappa}c(\omega)-c_{in}(\omega)$, we can find the fluctuations $\delta c_{out}(\omega)$ of the output field. For the purpose of Fig. (\ref{Fig2}) we define the output field as $\tilde{c}_{out}(\omega)=c_{out}(\omega)+c_{in}(\omega)$, then we find the result
\begin{equation}\label{10}
\delta\tilde{c}_{out}(\omega)=V(\omega)\xi(\omega)+E(\omega)c_{in}(\omega)+F(\omega)c_{in}^{\dag}(-\omega),
\end{equation}
in which
\begin{eqnarray}\label{11}
V(\omega)&=&\frac{\sqrt{2\kappa}gc_{s}\omega_{m}i}{d(\omega)}[\kappa-i(\omega+\Delta)],\nonumber\\
E(\omega)&=&\frac{2\kappa}{d(\omega)}\{2ig^{2}|c_{s}|^{2}\omega_{m}+(\omega_{m}^{2}-\omega^{2}-i\gamma_{m}\omega)\nonumber\\& &\times[\kappa-i(\omega+\Delta)]\},\nonumber\\
F(\omega)&=&\frac{4\kappa}{d(\omega)}\omega_{m}g^{2}c_{s}^{2}i,
\end{eqnarray}
where
\begin{equation}\label{12}
d(\omega)=-4\omega_{m}\Delta g^{2}|c_{s}|^{2}+(\omega_{m}^{2}-\omega^{2}-i\gamma_{m}\omega)[(\kappa-i\omega)^{2}+\Delta^{2}].
\end{equation}
The first term on the right-hand side of Eq. (\ref{10}) refers to the contribution of the thermal noise of the movable mirror, and the other two terms represent the contribution of the squeezed vacuum. To illustrate the meaning of the last two terms, let the squeezed vacuum be a single mode i.e. $c_{in}(t)=Ce^{-i(\omega_{p}-\omega_{c})t}$, then $c_{in}(\omega)=2\pi C\delta(\omega-\omega_{p}+\omega_{c})$ and $c_{in}^{\dag}(-\omega)=2\pi C^{\dag}\delta(\omega+\omega_{p}-\omega_{c})$. Thus the fluctuations of the output field $\delta\tilde{c}_{out}(t)=\frac{1}{2\pi}\int^{+\infty}_{-\infty}V(\omega)\xi(\omega)e^{-i\omega t}d\omega+CE(\omega_{p}-\omega_{c})e^{-i(\omega_{p}-\omega_{c})t}+C^{\dag}F(\omega_{c}-\omega_{p})e^{-i(\omega_{c}-\omega_{p})t}$. Therefore, $E(\omega_{p}-\omega_{c})$ is the component at the probe frequency $\omega_{p}$ [which in rotating frame is $\omega_{p}-\omega_{c}$], and $F(\omega_{c}-\omega_{p})$ is the component at the new frequency $2\omega_{c}-\omega_{p}$ [which in rotating frame is $\omega_{c}-\omega_{p}$], due to the nonlinear interaction between the movable mirror and the cavity field.

By the aid of the correlation functions of the noise operators $c_{in}(\omega)$ and $\xi(\omega)$, and neglecting fast oscillating terms at frequency $\pm2\omega_{m}$, we obtain the homodyne spectrum $X(\omega)$ of the output field as measured by the set up of Fig. \ref{Fig2}
\begin{widetext}
\begin{eqnarray}\label{13}
X(\omega)&=&E(\omega+\omega_{m})E(-\omega+\omega_{m})\frac{M\Gamma^{2}}{\Gamma^{2}+\omega^{2}}+|E(\omega+\omega_{m})|^{2}\frac{N\Gamma^{2}}{\Gamma^{2}+\omega^{2}}\nonumber\\& &+E^{*}(-\omega+\omega_{m})E^{*}(\omega+\omega_{m})\frac{M\Gamma^{2}}{\Gamma^{2}+\omega^{2}}+|E(-\omega+\omega_{m})|^{2}\frac{N\Gamma^{2}}{\Gamma^{2}+\omega^{2}}\nonumber\\& &+|E(\omega+\omega_{m})|^{2}+|F(-\omega+\omega_{m})|^{2}\nonumber\\
& &+|V(\omega+\omega_{m})|^{2}2\gamma_{m}\frac{\omega+\omega_{m}}{\omega_{m}}\left\{1+\coth\left[\frac{\hbar(\omega+\omega_{m})}{2k_{B}T}\right]\right\}\nonumber\\ \nonumber\\& &+|V(-\omega+\omega_{m})|^{2}2\gamma_{m}\frac{\omega-\omega_{m}}{\omega_{m}}\left\{1+\coth\left[\frac{\hbar(\omega-\omega_{m})}{2k_{B}T}\right]\right\},
\end{eqnarray}
\end{widetext}
where the first four terms in Eq. (\ref{13}) originate from the squeezed vacuum, the next two terms not involving $N$ and $M$ are the contributions of the spontaneous emission of the input vacuum noise; the last two terms in Eq. (\ref{13}) result from the thermal noise of the movable mirror.

\section{EIT in the homodyne spectrum of the output quantized field}
After having derived the homodyne spectrum of the output field, we next examine it numerically to explore the EIT phenomenon in the homodyne spectrum of the output field. Since the original equations (\ref{2}) are nonlinear, these can have instabilities. Thus in the following, we work in the stable regime of the system. We first examine the frequency at which we expect transparency. This is $\omega=0$. For $N\approx M$,
\begin{widetext}
\begin{eqnarray}\label{14}
X(0)&=&N[E(\omega_{m})+E^{*}(\omega_{m})]^{2}+|E(\omega_{m})|^{2}+|F(\omega_{m})|^{2}+4|V(\omega_{m})|^{2}\gamma_{m}\coth\left[\frac{\hbar\omega_{m}}{2k_{B}T}\right].
\end{eqnarray}
\end{widetext}

We use the parameters from the experimental paper \cite{Weis} focusing on the EIT in the optomechanical system: the wavelength of the coupling field $\lambda=2\pi
c/\omega_{c}=775$ nm, the coupling constant $g=2\pi\times12\ \mathrm{GHz/nm}\sqrt{\hbar/(2m\omega_{m})}$, the mass of the movable mirror $m=20$ ng, the frequency of the movable mirror $\omega_{m}=2\pi\times51.8$ MHz, the cavity decay rate $\kappa=2\pi\times15$ MHz, $\kappa/\omega_{m}=0.289$, the mechanical damping rate $\gamma_{m}=2\pi\times41$ kHz, the mechanical quality factor $Q'=\omega_{m}/\gamma_{m}=1263$. In addition, we choose the linewidth of the squeezed vacuum $\Gamma=2\kappa$, and consider the resonant case $\Delta=\omega_{m}$.

For $N=10$ and $M=\sqrt{N(N+1)}\approx10$, $\wp=20$ mW, $T=20$ mK, the first term in Eq. (\ref{14}) which is the contribution of the squeezed vacuum is about $6.5\times10^{-4}$, the sum of the second and third terms in Eq. (\ref{14}) which are the contribution of the input vacuum noise is about 0.16, the last term arising from the thermal noise of the movable mirror is about 0.14. The contribution of the input quantum field in principle can be obtained by doing the experiment with and without the quantized field and by subtracting the data i.e. by studying $X(0)-X(0)|_{N=0}$. The squeezed field part in a sense exhibits perfect EIT. If $M=0$, i.e. the input quantized field is phase insensitive, then such a field leads to a term $2N|E(\omega_{m})|^{2}$ which is equal to 1.6 for the above mentioned parameters and hence there is no perfect EIT. The squeezed field changes $2N|E(\omega_{m})|^{2}$ to $N[E(\omega_{m})+E^{*}(\omega_{m})]^{2}$ and for the above parameters the number changes from 1.6 to $6.5\times10^{-4}$.

\begin{figure}[htp]
 \scalebox{0.8}{\includegraphics{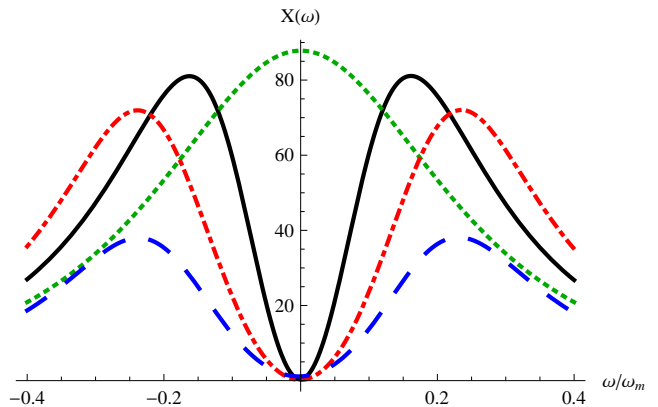}}
 \caption{\label{Fig3} Plots of the homodyne spectrum $X(\omega)$ as a function of $\omega/\omega_{m}$ for $N=5$ in the absence (dotted curve) and the presence (solid, dotdashed, and dashed curves) of the coupling field for the temperature of the environment $T=20$ mK. The solid curve is for $\wp=10$ mW and $M=\sqrt{N(N+1)}$, the dotdashed curve is for $\wp=20$ mW and $M=\sqrt{N(N+1)}$, the dashed curve is for $\wp=20$ mW and $M=0$.}
 \end{figure}

For $N=5$, $M=\sqrt{N(N+1)}$ and 0, $T=20$ mK, we plot the homodyne spectrum $X(\omega)$ of the output field as a function of the normalized frequency $\omega/\omega_{m}$ in the absence (dotted curve) and presence (solid, dotdashed, and dashed curves) of the coupling field in Fig. \ref{Fig3}. First let us look at the case that the input quantum field is phase dependent ($M=\sqrt{N(N+1)}$). In the absence of the coupling field, one can note that the homodyne spectrum of the output field has a Lorentzian line shape. However, in the presence of the coupling field at different power levels, the solid curve ($\wp=10$ mW and $M=\sqrt{N(N+1)}$) and the dotdashed curve ($\wp=20$ mW and $M=\sqrt{N(N+1)}$) exhibit the EIT dip, which is the result of the destructive interference between the squeezed vacuum and the scattering quantum field at the probe frequency $\omega_{p}$ generated by the interaction of the coupling field with the movable mirror. For $\wp=20$ mW and $M=\sqrt{N(N+1)}$, the minimum value of $X(\omega)$ is about 0.22. Moreover, the linewidth of the dip for $\wp=20$ mW is larger than that for $\wp=10$ mW due to power broadening. Generally the EIT dip has a contribution to its width which is proportional to the power of the coupling field. We indeed find that the width for $\wp=20$ mW is 0.26$\omega_{m}$, which is about twice the width for $\wp=10$ mW. If the input quantum field is phase independent ($M=0$) [the dashed curve], we can see that the maximum value of $X(\omega)$ for $\wp=20$ mW and $M=0$ is about half that for $\wp=20$ mW and $M=\sqrt{N(N+1)}$.

\begin{figure}[htp]
 \scalebox{0.8}{\includegraphics{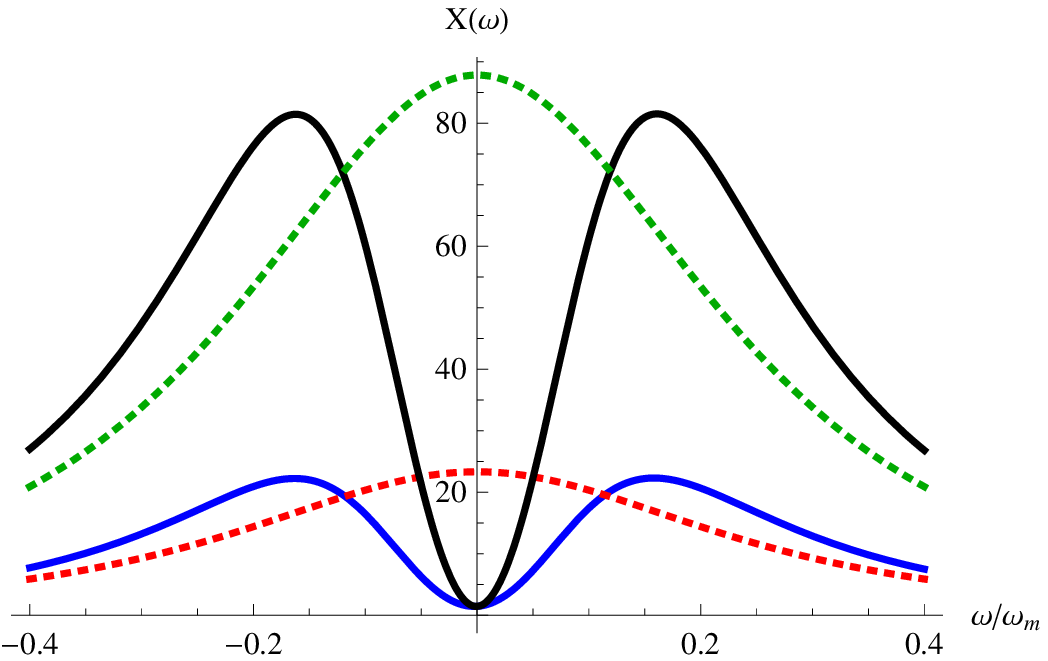}}
 \caption{\label{Fig4} Plots of the homodyne spectrum $X(\omega)$ as a function of $\omega/\omega_{m}$ for different values of the parameter $N$ and $M=\sqrt{N(N+1)}$ in the absence (dotted curves) and the presence (solid curves) of the coupling field with power $\wp=10$ mW and the temperature of the environment $T=100$ mK. The upper two curves are for $N=5$, the lower two curves are for $N=1$.}
 \end{figure}

Next we increase the temperature to 100 mK. The figure \ref{Fig4} displays the homodyne spectrum $X(\omega)$ of the output field against the normalized frequency $\omega/\omega_{m}$ in the absence (dotted curves) and presence (solid curves) of the coupling field for $N=1, 5$ and $M=\sqrt{N(N+1)}$. In the presence of the coupling field ($\wp=10$ mW), it is seen that the EIT dip still appears in the homodyne spectrum of the output field for $N=1$ and 5. Note that the two dips almost have the same minimum values (about 1.43) and the same linewidth (about 0.15$\omega_{m}$). Hence the temperature of the environment is not detrimental to the EIT behavior.

\begin{figure}[htp]
 \scalebox{0.8}{\includegraphics{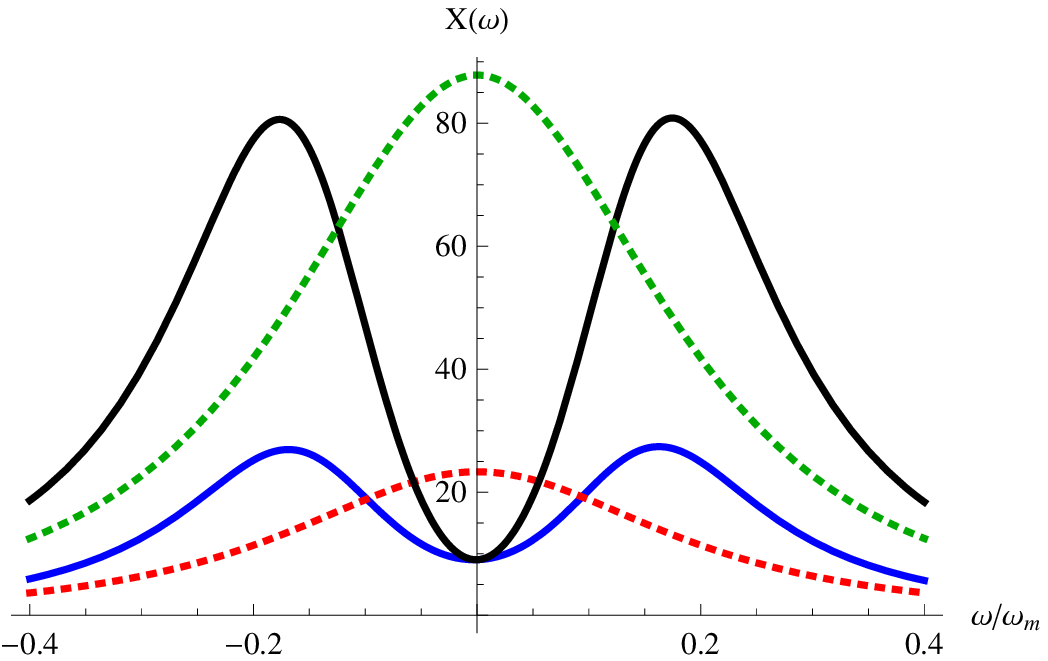}}
 \caption{\label{Fig5} Plots of the homodyne spectrum $X(\omega)$ as a function of $\omega/\omega_{m}$ for different values of the parameter $N$ and $M=\sqrt{N(N+1)}$ in the absence (dotted curves) and the presence (solid curves) of the coupling field with power $\wp=10$ mW and the temperature of the environment $T=100$ mK. The upper two curves are for $N=5$, the lower two curves are for $N=1$.}
 \end{figure}

The effects discussed above occur under wide range of parameters. We demonstrate this by using the experimental parameters \cite{Aspelmeyer}: $\lambda=2\pi
c/\omega_{c}=1064$ nm, $L=25$ mm, $g\approx2\pi\times11.28\ \mathrm{MHz/nm}\sqrt{\hbar/(2m\omega_{m})}$, $m=145$ ng, $\omega_m=2\pi\times947$ kHz, $\kappa=2\pi\times215$ kHz, $\kappa/\omega_{m}=0.227$, $\gamma_{m}=2\pi\times141$ Hz, $Q^{\prime}=\omega_{m}/\gamma_{m}=6700$. The values for parameters $T$, $\wp$, $N$, $M$, $\Gamma$, and $\Delta$ are the same as those in Fig. \ref{Fig4}. Shown in Fig. \ref{Fig5} is the homodyne spectrum $X(\omega)$ of the output field as the normalized frequency $\omega/\omega_{m}$ is varied for $T=100$ mK and $\wp=0, 10$ mW. Note that the EIT exists for $N=1$ and 5 in the presence of the coupling field. The linewidth of the dip for $N=5$ is about $0.2\omega_{m}$, and as expected gets broadened due to power. We have further studied the effect of temperature and we find that there is rather weak dependence of the EIT curves on temperature.  Therefore, current optomechanical designs can be used to realize quantum optical memory at single photon level.

\section{Conclusions}
In conclusion, we have demonstrated EIT using quantum fields in optomechanical systems under a wide range of conditions. For squeezed quantum fields we obtained perfect EIT. The EIT gets degraded in phase insensitive quantum fields. We have shown that even temperature is not critical for observations of EIT. The results get be generalized to optomechanical systems working on the reactive coupling \cite{Li,Sumei1,Sumei2}. Our work suggest that optomechanical systems could be used as elements for quantum memory.

We gratefully acknowledge support from the NSF Grant No. PHYS 0653494.

\end{document}